\documentclass[sigconf, nonacm]{acmart}

\usepackage{multirow}
\usepackage{makecell}
\usepackage{listings}
\usepackage{algorithm}
\usepackage{algpseudocode}
\usepackage{makecell}
\usepackage{booktabs}
\usepackage{multirow}
\usepackage[most]{tcolorbox}
\usepackage{array}
\usepackage[table]{xcolor}
\usepackage{geometry}
\usepackage{xcolor}
\usepackage{bbding}
\usepackage{changepage}
\usepackage{paracol}
\newcommand{\hl}[2][yellow]{\begingroup\setlength\fboxsep{1pt}\colorbox{#1}{#2}\endgroup}

\lstdefinestyle{mystyle}{
    backgroundcolor=\color{gray!10},   
    commentstyle=\color{black},
    keywordstyle=\color{blue},
    morekeywords={Execution, from, sqlite3},
    numberstyle=\tiny\color{gray},
    stringstyle=\color{red},
    basicstyle=\ttfamily\footnotesize,
    breakatwhitespace=false,         
    breaklines=true,                 
    captionpos=b,                    
    keepspaces=true,                 
    showspaces=false,                
    showstringspaces=false,
    showtabs=false,                  
    tabsize=2
}

\newcommand\vldbdoi{XX.XX/XXX.XX}
\newcommand\vldbpages{XXX-XXX}
\newcommand\vldbvolume{14}
\newcommand\vldbissue{1}
\newcommand\vldbyear{2020}
\newcommand\vldbauthors{\authors}
\newcommand\vldbtitle{\shorttitle} 
\newcommand\vldbavailabilityurl{https://github.com/VeyC/MCI-SQL}
\newcommand\vldbpagestyle{plain} 

\newcommand{\micsql}{MCI-SQL}



\begin{document}
\title{\micsql: Text-to-SQL with Metadata-Complete Context and Intermediate Correction}
\settopmatter{authorsperrow=4}

\author{Qin Wang}
\affiliation{
  \institution{Hunan University}
  \city{Hunan}
  \country{China}
}
\email{qinwang@hnu.edu.cn}

\author{Youhuan Li}
\authornote{Corresponding author.}
\affiliation{
  \institution{Hunan University}
  \city{Hunan}
  \country{China}
}
\email{liyouhuan@hnu.edu.cn}

\author{Suixi Lin}
\affiliation{
  \institution{Hunan University}
  \city{Hunan}
  \country{China}
}
\email{linsuixi@hnu.edu.cn}

\author{Zhou Tang}
\affiliation{
  \institution{Hunan University}
  \city{Hunan}
  \country{China}
}
\email{ztang@hnu.edu.cn}

\author{Kenli Li}
\affiliation{
  \institution{Hunan University}
  \city{Hunan}
  \country{China}
}
\email{lkl@hnu.edu.cn}

\author{Peng Peng}
\affiliation{
  \institution{Hunan University}
  \city{Hunan}
  \country{China}
}
\email{hnu16pp@hnu.edu.cn}

\author{Quanqing Xu}
\affiliation{
  \institution{OceanBase, Ant Group}
  \city{Zhejiang}
  \country{China}
}
\email{xuquanqing.xqq@oceanbase.com}

\author{Chuanhui Yang}
\affiliation{
  \institution{OceanBase, Ant Group}
  \city{Zhejiang}
  \country{China}
}
\email{rizhao.ych@oceanbase.com}

\begin{abstract}
Text-to-SQL aims to translate natural language queries into SQL statements. 
Existing methods typically follow a pipeline of pre-processing, schema linking, candidate SQL generation, SQL alignment, and target SQL selection.
However, these methods face significant challenges. 
First, they often struggle with column filtering during schema linking due to difficulties in comprehending raw metadata.
Also, the candidate SQL generation process often suffers from reasoning errors, which limits accuracy improvements.
To address these limitations, we propose a framework, called \micsql, to efficiently and precisely generate SQL queries.
Specifically, we assign metadata-complete contexts to each column, which significantly improves the accuracy of column filtering for schema linking.
Also, for candidate SQL generation, we propose an intermediate correction mechanism that validates SQL queries and revises errors in a timely way.
Moreover, we also propose effective optimizations in subsequent SQL alignment and selection phases, which further enhance the performance.
Experiments on the widely-used BIRD benchmark show that \micsql\ achieves execution accuracy of 74.45\% on the development set and 76.41\% on the test set, surpassing current published state-of-the-art results.
In addition, we manually identify and correct 412 samples in the BIRD
 dataset, forming a new version named BIRD-clear, which is released together with our code on GitHub.
We also evaluate our methods on BIRD-clear and find that \micsql\ outperforms baselines by 8.47 percentage points in execution accuracy, further demonstrating the effectiveness and reliability of our framework.
\end{abstract}

\maketitle

\pagestyle{\vldbpagestyle}
\begingroup\small\noindent\raggedright\textbf{PVLDB Reference Format:}\\
\vldbauthors. \vldbtitle. PVLDB, \vldbvolume(\vldbissue): \vldbpages, \vldbyear.\\
\href{https://doi.org/\vldbdoi}{doi:\vldbdoi}
\endgroup
\begingroup
\renewcommand\thefootnote{}\footnote{\noindent
This work is licensed under the Creative Commons BY-NC-ND 4.0 International License. Visit \url{https://creativecommons.org/licenses/by-nc-nd/4.0/} to view a copy of this license. For any use beyond those covered by this license, obtain permission by emailing \href{mailto:info@vldb.org}{info@vldb.org}. Copyright is held by the owner/author(s). Publication rights licensed to the VLDB Endowment. \\
\raggedright Proceedings of the VLDB Endowment, Vol. \vldbvolume, No. \vldbissue\ %
ISSN 2150-8097. \\
\href{https://doi.org/\vldbdoi}{doi:\vldbdoi} \\
}\addtocounter{footnote}{-1}\endgroup

\ifdefempty{\vldbavailabilityurl}{}{
\vspace{.3cm}
\begingroup\small\noindent\raggedright\textbf{PVLDB Artifact Availability:}\\
The source code, data, and/or other artifacts have been made available at \url{\vldbavailabilityurl}.
\endgroup
}

\section{Introduction}
Text2SQL is to convert natural language queries into corresponding SQL queries that can be executed on a relational database~\cite{yu2018spider, zhong2017seq2sql, bogin2019representing, yu2018typesql, wang2022proton, gu2023few}.
This task enables non-experts to query databases without  knowledge of SQL syntax.
\begin{figure}[!t]
	\centering
	\includegraphics[width=\linewidth]{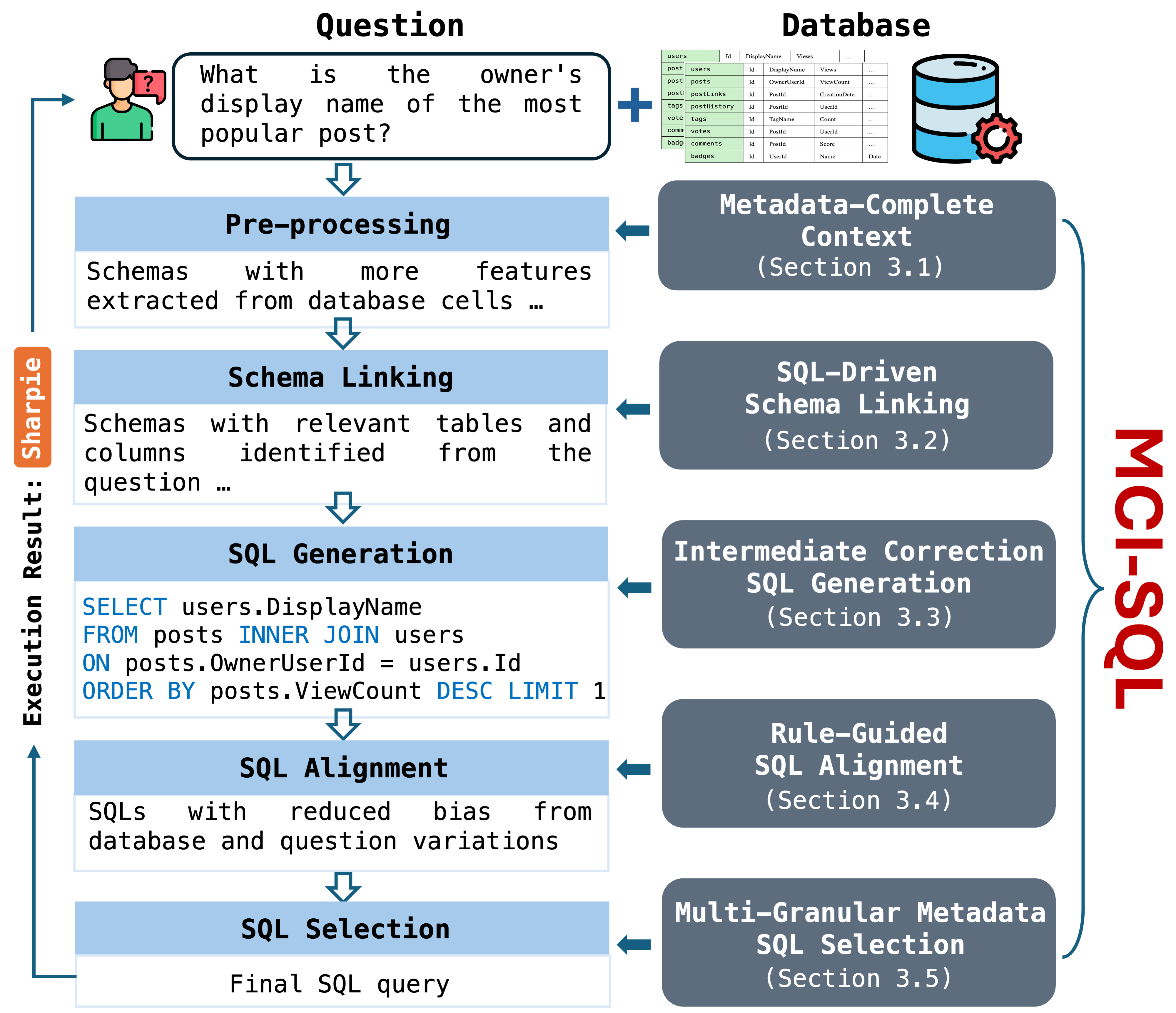}
	\caption{Overview of the prevalent LLM-based Text2SQL workflow and our proposed \micsql\ operations at each phase.} 
\label{pic:motivation}
\end{figure} 

Numerous studies have investigated Text2SQL~\cite{yin2016neural, yu2018typesql, guo2019towards, he2019x, kim2020natural, wang2022proton, gu2023few, wang2025mac, xie2025opensearch}. 
Recent advances leverage prompt engineering with large language models (LLMs) to further improve performance on the Text2SQL task.
These approaches typically decompose the task into a multi-stage process, where each stage is responsible for a distinct sub-task~\cite{gu2023few, xie2025opensearch, lee2025mcs, li2023resdsql, pourreza2024chase, wang2025mac}.
Figure~\ref{pic:motivation} presents a basic framework for prompt-based methods consisting of five main phases: (1) pre-processing to ensure model-ready database schemas; (2) schema linking for question-related column filtering; (3) candidate SQL generation over selected columns; (4) aligning generated SQL queries with preferences; and (5) selecting the final SQL as the output.
These methods offer superior generalization capabilities without the need for costly parameter updates. However, they still suffer from a performance bottleneck in generating correct SQL queries.
%
Specifically, prior work~\cite{pourreza2023din, gao2023text, xie2024decomposition, qu2024before, xie2025opensearch} makes limited use of database metadata, leading to a marked gap between LLMs and domain experts in column filtering during schema linking. Closing this gap is critical to improving model performance. 
Also, existing methods lack the ability to correct errors in a timely manner during SQL generation~\cite{pourreza2023din, wang2025mac, wang2025agentar}. They typically employ a separate post-hoc correction module to refine generated SQL queries, thereby overlooking reasoning details within the SQL generation process.

To this end, we propose an LLM-based workflow framework named \micsql.
As shown in Figure~\ref{pic:motivation}, in the pre-processing phase, we generate a complete metadata context for each column to help LLMs better understand databases.
Intuitively, LLMs require sufficiently rich metadata to understand individual columns for precise column filtering in the database schema. 
Therefore, we aggregate column-level, inter-column, and table-level metadata for each column. Then, based on this metadata, we generate natural language descriptions to serve as the metadata-complete context for the corresponding columns.
%
%
In the SQL generation phase, we propose an intermediate correction mechanism to validate SQL queries and revise them in time. 
Specifically, we decompose the question into sub-questions and convert them into subqueries.
Then, we validate and correct each subquery with execution feedback before assembling them into the final SQL.
This mechanism efficiently integrates the  processes of generation and correction, facilitating the timely correction of errors rather than addressing them at the final output, thereby substantially improving the success rate of error resolution.
%
Additionally, we incorporate specific optimizations in subsequent phases. 
To ensure consistency across successive phases, we apply a rule-generated alignment strategy to safeguard essential SQL components from unintended modifications.
We further design a pluggable SQL alignment phase and distinguish it into function alignment and output alignment to standardize SQL formulation.
For the SQL selection phase, we utilize multi-granular metadata to generate candidates and select the final SQL via majority voting.

We evaluate \micsql\ on the widely-used BIRD~\cite{li2023can} benchmark and achieve execution accuracy (EX) of 74.45\% on the development set and 76.41\% on the test set, surpassing published state-of-the-art methods.
Also, our method significantly improves the accuracy of single-SQL generation, thereby achieving an optimal balance between precision and efficiency.
Furthermore, we identify certain annotation errors and biases within the BIRD development set during evaluation, which involve $412$ samples and may lead to inaccurate assessments of model capabilities. 
Therefore, we manually correct these errors, and release a new version named BIRD-clear. 
We refine the dataset through a rigorous three-step process: individual annotation, cross review, and final inspection. 
Experiments on BIRD-clear further confirm the effectiveness of our method against existing open-source baselines.

In summary, our main contributions are as follows.
\begin{itemize}
    \item We propose \micsql\ to automatically generate metadata-complete context, which bridges the understanding gap between LLMs and databases.
    \item We introduce an intermediate correction mechanism that utilizes execution feedback to revise SQL queries in a timely way during the SQL generation process.
    \item We construct BIRD-clear by correcting 412 annotation errors in the original BIRD development set through a rigorous manual review.
    \item Experimental results demonstrate the superior performance of \micsql\ on both the original BIRD and BIRD-clear, validating the efficacy of our proposed framework.
\end{itemize}

\section{Preliminaries}
\subsection{Problem Definition}
\begin{definition}[Text2SQL] \label{def:text2sql}
Given a textual question and a database comprising its schema and cell values, Text2SQL aims to generate a SQL query whose execution on the database yields the correct answer to the question.
\end{definition}
Figure~\ref{pic:motivation} shows an example of the \textit{Codebase Community} database in the BIRD benchmark. Given the question:

\begin{center}
\itshape What is the owner's display name of the most popular post? ,
\end{center}
the task is to generate a matching SQL query, e.g.,

\lstdefinestyle{plainSQL}{
    language=SQL,
    basicstyle=\ttfamily\small,  
    keywordstyle=\bfseries,      
    frame=none,                  
    backgroundcolor=\color{gray!10}, 
    xleftmargin=0em,             
    breaklines=true,
    columns=flexible,            
    keepspaces=true,
    showstringspaces=false,
}
\begin{lstlisting}[style=plainSQL]
    SELECT users.DisplayName
    FROM posts INNER JOIN users
    ON posts.OwnerUserId = users.Id
    ORDER BY posts.ViewCount DESC LIMIT 1
\end{lstlisting}
to retrieve the correct answer, i.e., \textbf{\textit{sharpie}}.

\subsection{Related Work}

\subsubsection{Deep Learning-Based Text2SQL} 
Text2SQL serves as a critical interface for natural language to database interaction. 
With recent advances in deep learning, Text2SQL has attracted extensive interest from both academia and industry. 
Before the emergence of LLMs, Text2SQL research mainly relies on Seq2Seq architectures~\cite{graves2012long, vaswani2017attention, bogin2019representing} that jointly encode natural language questions and database schemas and then generate SQL queries using the decoder, such as IRNet~\cite{jha2019irnet}, Seq2SQL~\cite{zhong2017seq2sql}, SQLNet~\cite{xu2017sqlnet}, Ryansql~\cite{choi2021ryansql}, and Resdsql~\cite{li2023resdsql}.
Following the release of open-source LLMs~\cite{du2022glm, touvron2023llama, hui2024qwen2}, the research has shifted to fine-tuning vanilla LLMs for Text2SQL tasks using large-scale data. Representative works include CodeS~\cite{li2024codes}, Omni-SQL~\cite{li2025omnisql}, and Xiyan-SQL~\cite{liu2025xiyan}. To further improve model performance, Arctic-Text2SQL-R1~\cite{yao2025arctic} employs reinforcement learning for post-training. This encourages the model to generate SQL queries that are both syntactically correct and executionally accurate.
Another line of research explores prompt engineering based on powerful closed-source models, such as GPT-4~\cite{achiam2023gpt}. This approach leverages the strong in-context learning and reasoning capabilities of LLMs to generate SQL via carefully designed prompts.
Specifically, DAIL-SQL~\cite{gao2023text}, DEA-SQL~\cite{xie2024decomposition}, and PURPLE~\cite{ren2024purple} focus on efficient demonstration selection strategies for few-shot learning. Additionally, methods like DIN-SQL~\cite{pourreza2023din}, CoE-SQL~\cite{zhang2024coe}, and ACT-SQL~\cite{zhang2023act} decompose the generation process into multiple sub-steps, either manually or automatically, to handle complex queries. Furthermore, DEA-SQL~\cite{xie2024decomposition}, MAC-SQL~\cite{wang2025mac}, and OpenSearch-SQL~\cite{xie2025opensearch} utilize multi-agent systems to break down the process into more fine-grained stages.

\subsubsection{Metadata-Based Context for Text2SQL}
Prompts for LLMs require a clear and comprehensive description of all information needed to answer the question. Since databases contain vast amounts of data that cannot be directly fed into LLMs, most existing works~\cite{pourreza2023din, dong2023c3, liu2023comprehensive, xie2024decomposition, xie2025opensearch} rely solely on schema metadata to represent the database structure.
Recently, some studies have attempted to enrich the database metadata. For example, TA-SQL~\cite{qu2024before} generates functional descriptions of each column. Works such as OpenSQL~\cite{chen2024open} and ACT-SQL~\cite{zhang2023act} include sample values from the database to help LLMs better understand the data. LTMP-DA-GP~\cite{arora2023adapt} adds value ranges for numerical columns, while SQLfuse~\cite{zhang2024sqlfuse} incorporates one-to-many relationships between tables. Additionally,~\cite{shkapenyuk2025automatic} provides fine-grained specifications of the storage format for each column.
Although these methods aim to bridge the gap between LLMs and databases, most of them are largely heuristic and fail to construct a complete metadata system, which leads to persistent challenges in accurate database comprehension.

\subsubsection{Correction Module for Text2SQL}
To improve SQL accuracy, most studies introduce reflection and correction modules. These methods typically utilize erroneous SQL and execution error information as prompts for an LLM to regenerate the SQL query. For instance, DIN-SQL~\cite{pourreza2023din} uses human-written guidelines to correct generated SQL queries. 
MAC-SQL~\cite{wang2025mac} and Agentar-Scale-SQL~\cite{wang2025agentar} employ a refinement component that uses execution errors as signals for correction. Similarly, RSL-SQL~\cite{chen2023teaching} and OpenSearch-SQL~\cite{xie2025opensearch} iterate the generation process until a non-empty result is obtained or a retry limit is reached. DEA-SQL~\cite{xie2024decomposition} analyzes common errors in field matching and syntax, then designs specific prompts to correct them.
Unlike these approaches, we integrate the correction module directly into the generation phase. This approach bridges the information gap between generation and correction to achieve superior performance.

\section{Method}
As illustrated in Figure~\ref{pic:motivation}, the proposed \micsql\ consists of five phases: metadata-complete context construction, schema linking, SQL generation, SQL alignment, and SQL selection. These phases run in a workflow manner and ultimately produce a final SQL query, which returns the result required by the user upon execution.

\subsection{Metadata-Complete Context}
The core challenge in Text2SQL tasks is to enable LLMs to understand databases and map user intentions to SQL queries.
Human comprehension of databases relies on rich prior knowledge and business context, whereas LLMs inherently possess only general syntactic knowledge. 
For example, consider a column named \textit{money} stored as an integer in a database. Humans familiar with the business logic can intuitively discern its currency type (e.g., US dollar or Canadian dollar) and whether the value is stored in cents or in dollars.
Furthermore, humans can distinguish subtle nuances between this column and similar ones (e.g., \textit{balance}) and infer the overall data distribution by inspecting the table contents. 
However, database schemas rarely enforce explicit specifications for such information, rendering it invisible to the LLM and resulting in a semantic gap regarding the column's true meaning.
Therefore, to bridge this gap, we construct metadata-complete contexts for databases, enabling LLMs to better understand the underlying data and generate accurate queries.
\begin{table*}[htpb]
    \centering
    \caption{Taxonomy of database metadata.}
    \small
    \setlength{\tabcolsep}{0.1cm}
    \begin{tabular}{|p{1.4cm}|p{1.3cm}p{3.8cm}|p{1.3cm}p{3.8cm}|p{1.4cm}p{3.4cm}|}
        \hline
        \multirow{2}{*}{\textbf{Category}} & \multicolumn{2}{c|}{\textbf{Column-Level}} & \multicolumn{2}{c|}{\textbf{Inter-Column-Level}} & \multicolumn{2}{c|}{\textbf{Table-Level}} \\
        \cline{2-7}
        & \textbf{Metadata} & \textbf{Description} & \textbf{Metadata} & \textbf{Description} & \textbf{Metadata} & \textbf{Description} \\
        \hline
        \textbf{Structure}
        & \textit{\makecell[l]{name \\ data type}}
        & \makecell[l]{Name of columns \\ Data type of columns}
        & \textit{\makecell[l]{foreign key}}
        & \makecell[l]{links between two columns}
        & \textit{\makecell[l]{name \\ primary key}}
        & \makecell[l]{Name of tables \\ primary key of tables} \\
        \hline
        \textbf{Semantics} 
        & \textit{\makecell[l]{description}} 
        & \makecell[l]{Functional meaning of columns}
        & \textit{similarity}
        & \makecell[l]{semantic similarity of columns}
        & \textit{\makecell[l]{description}}
        & \makecell[l]{Functional meaning of tables} \\
        \hline
        \textbf{Statistics}
        & \textit{\makecell[l]{range \\ pattern \\ example \\ null value \\ distinct \\ size \\ \\}}
        & \makecell[l]{Minimum and maximum values  \\ Value patterns (e.g., Aa9) \\ Example of values \\ Number of null values \\ Number of distinct values  \\ Minimum number of digits/ \\ characters}
        & \textit{\makecell[l]{correlation \\ \quad \\ distribution \\ similarity \\ dependency \\ \quad }}
        & \makecell[l]{Association between columns,\\  e.g., \textit{age} and \textit{salary} in \textit{User} table \\ Overlap in value distribution sha-\\ pes between two columns \\ Determinant relationships be- \\tween columns}
        & \textit{\makecell[l]{rows \\ columns}}
        & \makecell[l]{Number of rows \\ Number of columns} \\
        \hline
    \end{tabular}
    \label{table:metadata_summary}
\end{table*}

\subsubsection{Principles of Constructing Metadata-Complete Context}
Metadata refers to descriptive information about the data. In traditional database systems, metadata typically includes table structures, column names, data types, and primary-foreign key constraints defined in the schema. 
Such metadata primarily captures the logical structure and storage organization of the database.
While this structural information is essential, it provides only a limited view of the database. 
Therefore, we propose the principles of constructing metadata-complete contexts for databases. 
Formally, we define a comprehensive metadata space, denoted as $\Sigma$. For any given database element $D$, we aggregate all associated metadata into a set $S(\Sigma, D)$. 
Subsequently, $S(\Sigma, D)$ is transformed into a natural language description, denoted as $NL(\Sigma, D)$, which serves as the context for element $D$, thereby facilitating a comprehensive understanding by LLMs.
For instance, the metadata-complete context $NL(\Sigma, \text{`ViewCount'})$ for the \textit{ViewCount} column in the \textit{Codebase Community} database can be formulated as follows:
\begin{adjustwidth}{0cm}{0cm}
\begin{tcolorbox}[
    colback=gray!10,  
    colframe=black,        
    boxrule=0.4pt,         
    left=5mm, right=5mm,   
    top=1mm, bottom=1mm,   
    sharp corners           
]
\itshape
The `ViewCount' column in the `posts' table is an integer that represents the total number of times a post has been viewed. A higher view count means the post has higher popularity. The typical examples are 1,278 and 8,198. It ranges from 1 to 175,495. Multiple posts with different IDs may have the same view count ...
\end{tcolorbox}
\end{adjustwidth}

In this paper, to construct metadata-complete contexts, we systematically model metadata along two dimensions: (1) heterogeneous information sources, and (2) varying levels of granularity.  
From the perspective of information sources, database metadata can be divided into three categories: 
\begin{itemize}
    \item \textbf{Structure metadata}: definitions from the database schema.
    \item \textbf{Semantics metadata}: descriptions of the meanings of data.
    \item \textbf{Statistics metadata}: information derived from data instances.
\end{itemize}
In terms of granularity, database metadata can also be classified into three categories: 
\begin{itemize}
    \item \textbf{Column-level metadata} describes the features of columns.
    \item \textbf{Inter-column metadata} focuses on relationships between different columns.
    \item \textbf{Table-level metadata} describes the function of tables.
\end{itemize}
The metadata space constructed according to this taxonomy can provide high-coverage information to an LLM without direct access to raw data instances. Theoretically, it can capture all key elements required for SQL generation.
However, a database may involve a vast variety of metadata, making exhaustive enumeration neither feasible nor necessary for defining a taxonomy. Therefore, as shown in Table~\ref{table:metadata_summary}, we present representative metadata to illustrate each category rather than listing all possibilities.
The following subsection provides details on how we extract and organize metadata based on these granularity levels.

\subsubsection{Column-Level Metadata}
Extensive research efforts have been dedicated to column-level metadata. 
Although we construct a detailed column metadata space, we do not utilize all of it during inference. 
Specifically, the selected column-level metadata includes column name, data type, description, range, pattern, and data example. 
For column descriptions, we directly leverage the semantic descriptions generated by TA-SQL~\cite{qu2024before}. 
For value ranges, we use the SQL execution engine to calculate the minimum and maximum values for all numerical columns. 
Regarding data examples, we use BM25~\cite{robertson1994some} to dynamically retrieve values that are most relevant to the current question. 
For value patterns, we randomly sample 200 values from non-numerical columns and use an LLM to summarize their typical formatting patterns.
Finally, we integrate this enhanced information with the original column names and data types from the schema to construct the final context.
Note that this feature pruning is a temporary adaptation. We anticipate that as LLMs improve in handling complex contexts, all metadata will be utilized for deeper data comprehension.

\subsubsection{Inter-Column-Level Metadata}
We investigate inter-column metadata, which has been largely overlooked in existing studies.
We selectively apply inter-column metadata to address two core challenges in Text2SQL. 
One concerns semantically similar and duplicate columns, and the other involves dependencies between columns.
In large-scale databases, different tables often contain many columns that are semantically related but differ in subtle meanings. Additionally, to optimize performance, the same data may be repeatedly stored in multiple tables under different column names. These factors make the mapping from user questions to the correct columns highly sensitive and error-prone.
To address this challenge, unlike previous work, we first identify a set of semantically related candidate columns and then perform a more fine-grained categorization into duplicate columns and similar columns. 
Duplicate columns refer to those whose data content is fully equivalent. In contrast, similar columns share semantic and statistical similarities, but their concrete data instances are not identical. 
Specifically, we first use a sentence encoder to retrieve semantically similar columns. We then leverage an LLM to identify join paths among these columns based on the database schema.
Subsequently, we construct SQL queries to verify value consistency. 
This process enables the model to better capture relationships among columns and improves the accuracy of column filtering.

Moreover, the cardinality relationships between columns play a critical role in SQL generation, which directly affects filtering conditions and join operations.
Therefore, we capture the cardinality relationships between different columns to enhance the model's understanding.
To this end, we use the TANE algorithm~\cite{huhtala1999tane} to mine functional dependencies by analyzing the value distribution of column pairs within the same table.
For example, for column $c_a$ and $c_b$, if $c_a$ functionally determines $c_b$ (i.e., $c_a \rightarrow c_b$) but not vice versa, each value in $c_a$ corresponds to a unique value in $c_b$, whereas a value in $c_b$ may map to multiple values in $c_a$. 
This indicates an \textit{N:1} cardinality mapping between them. Similarly, we can derive \textit{1:1} and \textit{N:M} relationships.
These cardinality constraints help the model generate accurate SQL queries by deciding whether to use \textit{DISTINCT}, \textit{GROUP BY}, nested queries, etc.

\subsubsection{Table-Level Metadata}
We aggregate column-level metadata to construct a comprehensive table-level representation. Specifically, we employ LLMs to generate table descriptions based on schemas enriched with column-level metadata. These descriptions outline the table's function, identify key columns, and suggest potential usage scenarios.
This explicit definition of design intent and highlighting of core content assist the model in filtering out noise and focusing on critical information.

\subsection{SQL-Driven Schema Linking}
In this section, we extract columns that are relevant to the user question from the schema $schema_{m}$, which is enriched with the constructed metadata contexts, as not all columns contribute to the final answer. 
We leverage the LLM to generate a draft SQL query, denoted as $SQL_{d}$, and then parse it to identify the referenced columns, thereby forming a candidate column set $\mathcal{C} = \{c \mid c \in SQL_{d}\}$.
This choice is motivated by the fact that LLMs are better at generating SQL queries than at explicitly identifying relevant columns~\cite{li2024pet, shkapenyuk2025automatic}.
We require the model to avoid using explicit aliases. This constraint improves the executability of $SQL_{d}$ and increases the success rate of column extraction.
Furthermore, we augment the set $\mathcal{C}$ by incorporating related columns identified through inter-column metadata. 
Specifically, for each column $c_i \in \mathcal{C}$, we obtain its similar columns (denoted as $\mathcal{C}_{s}(c_i)$) and duplicate columns (denoted as $\mathcal{C}_{d}(c_i)$). The candidate column set is then updated as:
\begin{equation}
    \mathcal{C} \leftarrow \mathcal{C} \cup \bigcup_{c_i \in \mathcal{C}} \left( \mathcal{C}_{s}(c_i) \cup \mathcal{C}_{d}(c_i) \right)
\end{equation}
Finally, this augmented set $\mathcal{C}$ constitutes the filtered schema, formally denoted as ${schema}_{f}$.

\begin{figure*}[!t]
	\centering
	\includegraphics[width=0.9\linewidth]{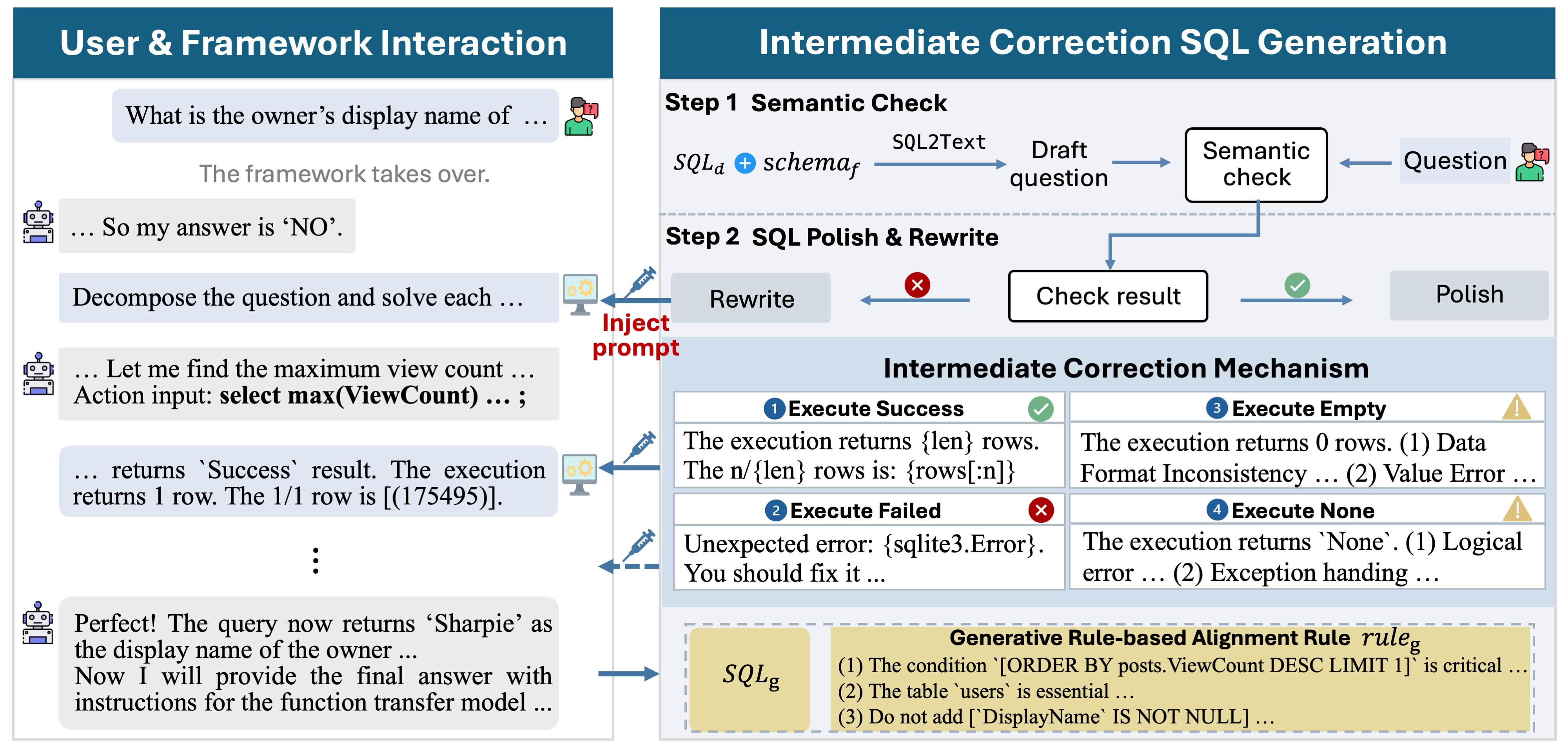}
	\caption {The framework of intermediate correction SQL generation phase.} 
\label{pic:method}
\end{figure*}
\subsection{Intermediate Correction SQL Generation}
In this section, we generate the SQL query using the question, the filtered schema $schema_{f}$, the draft query $SQL_{d}$, and similar cases retrieved via MQS~\cite{guo2023case}.
We instruct the LLM to follow a predefined reasoning process:
(1) Semantic check: The LLM evaluates whether $SQL_{d}$ aligns with the user’s intent by analyzing the semantics of the draft query, resulting in a binary decision (yes or no). 
(2) SQL generation: If the intent is met, the LLM performs fine-grained polishing on $SQL_{d}$; otherwise, it rewrites the SQL query. Ultimately, this process produces an SQL query, denoted as $SQL_{g}$.

\subsubsection{Adaptive Prompt Chaining}
We employ adaptive prompt chaining to enable LLMs to perform structured and controllable reasoning for SQL generation. 
%
Different from prior work, we assign both SQL generation and correction to a single LLM to preserve complete semantic information and implicit reasoning states. However, this unified design introduces new challenges. In long-chain reasoning scenarios, it becomes difficult to consistently maintain reliable instruction following. 
To address this issue, instead of relying on a single static prompt, our method adaptively injects pre-designed prompts that are highly relevant to the current subtask based on the model's reasoning state. This design is inspired by the observation that LLMs are often more sensitive to tokens near the beginning and end of the input sequence~\cite{hsieh2024found}. We place targeted prompts at key states while preserving the full reasoning context. 
In this way, the model is encouraged to refocus on the next objective, thereby reducing instruction drift and information loss.

Specifically, as illustrated in Figure~\ref{pic:method}, we first present the model with the overall task objective and predefined reasoning steps. 
Upon completing the semantic check, we interrupt the inference process based on predefined prompts and specific keywords.
The proposed framework then parses the checking result and determines which subtask prompt should be injected next. 
Throughout the entire generation process, prompts are always automatically injected according to the current reasoning state.
Listing~\ref{lst:sql_generation} shows the detailed prompt for the SQL generation phase.
\definecolor{myblue}{RGB}{0,0,255}
\begin{lstlisting}[style=mystyle, caption={The prompt for SQL generation.}, label={lst:sql_generation}, escapeinside={(*@}{@*)}]
# Goal: Follow the STEP, generate an executable SQL that fully satisfies the user question. 
# STEP:
1. Analyze the Draft SQL and explain what this query is intended to do. Determine whether it is suitable to answer the question. Answer only "YES" or "NO" after careful thinking.
2.1. If your answer is "NO", rewrite the executable SQL query to correctly answers the question. 
2.2. If your answer is "YES", check and correct any minor errors existing in Draft SQL.
** VERY IMPORTANT: After completing step 1, STOP generation. Wait for the system to provide additional instructions.**
# Database Schema: {(*@\textcolor{myblue}{$schema_f$}@*)}
# Output Format: {(*@\textcolor{myblue}{output format}@*)}
Question: {(*@\textcolor{myblue}{question}@*)}
Draft SQL: {(*@\textcolor{myblue}{$SQL_d$}@*)}
\end{lstlisting}

\subsubsection{Polishing \& Rewriting}
After semantic validation, we select one of two SQL generation paths based on the validation result, namely polishing and rewriting.
When the semantic validation result is `yes', the framework enters the polishing branch. 
The goal of this stage is to refine the draft SQL through fine-grained optimization and constraint checking. This stage mainly focuses on correcting minor errors that frequently occur in SQL generation, such as the use of \textit{DISTINCT} and the handling of \textit{NULL} values. During this process, the model can access column dependencies through a metadata tool to determine whether operations like \textit{DISTINCT} are required. In addition, the model can validate the generated SQL by executing it via an SQL engine. If the execution result does not meet the model's expectations, an intermediate correction mechanism is triggered to revise the SQL until the result is satisfactory.
When the semantic validation result is `no', the framework enters the rewriting branch. In this stage, the model regenerates the SQL query to better satisfy the user's intent. Specifically, the model first decomposes the user question into several sub-questions and generates a corresponding subquery for each sub-question. The final SQL query is formed by progressively combining the subqueries. If a subquery fails to meet the model's expectations, the framework timely activates the intermediate correction mechanism. This design ensures that each sub-question is correctly solved before proceeding to the next step, which enables timely detection and correction of errors.

\subsubsection{Intermediate Correction Mechanism}
Constructing correct SQL queries typically involves a cycle of execution, result analysis, and modification. 
Therefore, instead of performing direct generation and then correction, 
we introduce an intermediate correction mechanism that allows the model to correct the SQL queries during the generation process.
Specifically, we dynamically inject targeted instructions based on the current SQL execution state. We define four distinct states:

\textbf{(1) Execution Success:} The SQL query executes successfully without triggering the exception states described below. We truncate the result set to the first \textit{n} rows and return it to the model as feedback. 
Based on this observation, the model autonomously determines its subsequent actions. The instructions are shown in Listing~\ref{lst:exec_sucess}.
\begin{lstlisting}[style=mystyle, caption={The instruction for Execution Success.}, label={lst:exec_sucess}, escapeinside={(*@}{@*)}]
The SQL query {(*@\textcolor{myblue}{sql}@*)} returns `Success`.  
The execution returned {(*@\textcolor{myblue}{len(rows)}@*)} rows. 
The {(*@\textcolor{myblue}{n}@*)}/{(*@\textcolor{myblue}{len(rows)}@*)} rows are: {(*@\textcolor{myblue}{rows[:n]}@*)}
\end{lstlisting}

\textbf{(2) Execution None:} This state indicates that the SQL executes successfully but returns \textit{None} values. This serves as a strong signal of invalid logic. 
In this state, LLMs tend to filter out \textit{None} values (e.g., by adding \textit{IS NOT NULL}) to obtain valid results. 
However, this approach typically violates the user's intent unless the question explicitly contains filtering requirements.
Specifically, this anomaly can stem from schema linking bias, where the model selects an incorrect column from semantically similar candidates, such as \textit{Displayname} in the \textit{user} table versus \textit{OwnerDisplayName} in the \textit{post} table.
To address this, we inject instructions that discourage naive value filtering and guide the model to reconsider the column selection and explore an alternative reasoning path.
The detailed instructions are provided below.
\begin{lstlisting}[style=mystyle, caption={The instruction for Execution None.}, label={lst:exec_none}, escapeinside={(*@}{@*)}]
The SQL query {(*@\textcolor{myblue}{sql}@*)} returns `None`. 
You should consider whether there is a potential issue below, and adjust your answer to return valid results.
(1) **Logical error:** Following the SQL skeleton provided in the example, you should try another reasoning path.
(2) **Exception:** Do not introduce additional filters to exclude outliers in order to avoid returning `None` result, unless the question explicitly instructs you to do so.
\end{lstlisting} 

\textbf{(3) Execution Empty:} The SQL query returns an empty list or a count of zero. This result suggests that the generated SQL contains overly strict filters or incorrect join predicates. 
Without explicit guidance, the model tends to explore blindly within a vast hypothesis space, which leads to low efficiency. 
Therefore, we inject instructions to specifically correct filter conditions.
First, discrepancies in data storage formats may cause value mismatches. For instance, a query condition like \textit{CharterNum = `40'} may need to be converted to \textit{CharterNum = `0040'} to match the actual stored data.
In such cases, we prompt the model to re-examine the data pattern of the target column to align the format.
Second, for string columns, exact matching may result in retrieval failures. 
To resolve this, we instruct the model to relax constraints by adopting fuzzy matching, such as using the \textit{LIKE} operator instead of `='.
Finally, we advise the model to verify the join conditions to avoid generating empty results due to incorrectly connected columns. 
Listing~\ref{lst:exec_empty} presents the instructions for this state.
\begin{lstlisting}[style=mystyle, caption={The instruction for Execution Empty.}, label={lst:exec_empty}, escapeinside={(*@}{@*)}]
The SQL query {(*@\textcolor{myblue}{sql}@*)} returns `Empty`. 
You should consider whether there is a potential issue below, and adjust your answer to return valid results.
(1) Data Format Inconsistency: The values in the SQL query must be converted to the same format as the corresponding values in the database. 
(2) Value Mismatch: First, use case-insensitive fuzzy matching (e.g., `LOWER`,`LIKE`) to broaden the search and retrieve a subset of potential values. Then, within this subset, use a strict method (e.g., `=`) to identify the single correct value that best matches the user's intent.
(3) Domain Mismatch: Ensure that the joined columns or compared columns share the same semantic domain.
\end{lstlisting}

\textbf{(4) Execution Failure:} The SQL execution encounters an error. We return the specific error message back to the model to facilitate syntax debugging.
\begin{lstlisting}[style=mystyle, caption={The instruction for Execution Failure.}, label={lst:exec_fail}, escapeinside={(*@}{@*)}]
The SQL query {(*@\textcolor{myblue}{sql}@*)} returns `Failed`. 
You should fix it based on the message below.
Error Message: {(*@\textcolor{myblue}{sqlite3.OperationalError}@*)}
\end{lstlisting}

\subsubsection{Rule-Generated Alignment}
Workflow frameworks often suffer from limited inter-agent communication, leaving agents to operate on sub-tasks without shared context.
This limitation frequently leads to the recurrence of errors. 
While some existing studies~\cite{xie2025opensearch} employ alignment models to bridge agents, these approaches incur significant computational overhead. 
In contrast, we design a lightweight alignment mechanism to enforce consistency.
This module captures key insights from the tool-use and feedback loop during the SQL generation phase and encapsulates them as global rules for subsequent phases.
Specifically, we automatically generate rules $rule_g$ across three dimensions. 
As shown in Figure~\ref{pic:method}, first, condition rules identify critical filters whose removal leads to empty outputs. 
Second, table rules specify essential tables required for valid joins. 
Third, negative constraints highlight unnecessary conditions that should be pruned. 
These rules are propagated to downstream agents to guide exploration.

\subsection{Rule-Guided SQL Alignment}
This phase comprises function alignment and output alignment.
First, SQL queries that appear logically identical can have different syntax or implementations, potentially leading to inconsistent behaviors, such as \textit{DATE(`2026-01-01')} and \textit{CAST(`2026-01-01' AS DATE)}.
To mitigate this, we manually distill a set of function alignment rules. These rules map diverse function implementations to a standard form. Function alignment takes $rule_g$ and $SQL_g$ as input and produces $SQL_f$ as output.
The prompt is shown in Listing~\ref{lst:function_alignment}.
\begin{lstlisting}[style=mystyle, caption={The prompt for function alignment.}, label={lst:function_alignment}, mathescape=true, escapeinside={(*@}{@*)}]
# Goal: Your task is to perform a preference check on the given SQL query. You must strictly follow both the given rules and the check rules below, and convert the given SQL into a compliant, executable SQL query.
# Given Rules: {$\color{blue}{rule_g}$}
# Check Rules:
(1) Value Check Rule: Prohibit value checks (such as IS NOT NULL) unless the user explicitly requests them in the Question.
(2) Date Check Rule: For age calculation, use STRFTIME('%Y', time_now)-STRFTIME('%Y', Birthday). 
...
# Output Format: {(*@\textcolor{myblue}{output format}@*)}
# Given SQL: {$\color{blue}{SQL_g}$}
# Question: {$\color{blue}{\text{quesiton}}$}
\end{lstlisting}

Second, output alignment focuses on aligning the columns in the \textit{SELECT} clause. 
In pilot experiments, we observe that the model tends to return additional information. For example, when the user requests only the \textit{age} of students, the model might return both the student \textit{ID} and \textit{age}.
To improve precision, as shown in Listing~\ref{lst:output_alignment}, we employ a few-shot strategy that provides nine examples containing questions, correct SQL queries, incorrect SQL queries, and corresponding explanations. 
Based on these examples, the LLM adjusts the output columns and generates $SQL_o$.
\begin{lstlisting}[style=mystyle, caption={The prompt for output alignment.}, label={lst:output_alignment}, mathescape=true, escapeinside={(*@}{@*)}]
# Goal: Your task is to perform a column check on the given SQL query.
# STEP:
(1) Extract the explicit content that the user needs to return as the **minimum** requirement in the question. Omitting identifier columns is acceptable if not explicitly requested.
(2) Modify the SELECT clause in the SQL query to return only the requested content.
# Important Note:
**You are only allowed to delete, add, or reorder the selected columns. Other operations are strictly prohibited, even if the logic in the SQL might be incorrect.**
# Examples: {$\color{blue}{\text{examples}}$}  
# Output Format: {$\color{blue}{\text{output format}}$}
# Given SQL: {$\color{blue}{SQL_f}$}
# Question: {$\color{blue}{\text{quesiton}}$}
\end{lstlisting}
Importantly, these alignment rules do not alter the underlying logic of the SQL query. 
As a result, the alignment module is pluggable and can be easily removed or adapted for different scenarios. 

\subsection{Multi-Granular Metadata SQL Selection}
We generate multiple candidate SQL queries and select the final SQL $SQL_q$ using a multi-granular metadata-based voting strategy to improve execution accuracy. 
Specifically, we vary the metadata context during schema linking to obtain diverse candidate SQL queries. 
In the first setting, the model receives metadata-partial schemas, which include only column descriptions and representative values.
It generates $n$ SQL queries $SQL^p_S = \{q^p_1, q^p_2,...,q^p_n\}$, where $q^p_i$ denotes the SQL query generated after the SQL alignment phase. In the second setting, the model utilizes the metadata-complete schemas to generate $m$ SQL queries $SQL^c_S = \{q^c_1, q^c_2,...,q^c_m\}$. 
Consequently, we obtain a set of candidate SQL queries, denoted as $SQL_S=SQL^p_S \cup SQL^c_S$. We perform majority voting on the execution results of $SQL_S$ to determine the final SQL $SQL_q$.
Compared to prior work~\cite{wang2025agentar, xie2025opensearch, talaei2024chess}, our approach substantially reduces the number of candidate SQL queries and lowers generation cost.

\section{Experiments}
\subsection{Experimental Setup}
\subsubsection{Dataset}
We evaluate our method using Spider~\cite{yu2018spider} and BIRD~\cite{li2023can}, two widely adopted benchmark datasets for Text2SQL.
Spider covers 200 relational databases from various domains with relatively simple table structures. In total, the dataset includes 11,840 samples, with the training, development, and test sets containing 8,659 samples, 1,034 samples, and 2,147 samples, respectively.
BIRD is a larger dataset that closely reflects real-world application scenarios. In contrast to Spider, BIRD presents more challenging questions, more complex database schemas, and a substantial volume of real-world data.
These characteristics impose stricter demands on the model's semantic understanding and its capacity to generate complex SQL queries. The BIRD dataset comprises 9,428 training samples, 1,534 development samples, and 1,789 testing samples.

\subsubsection{Evaluation}
We adopt execution accuracy (EX) as our primary evaluation metric, which is the most widely used metric for Text2SQL. EX evaluates the prediction by executing the predicted SQL query on the corresponding database. It compares the execution result against that of the ground truth query. The prediction is considered correct only if the two execution results match exactly.

\subsubsection{Baseline}
We compare our proposed \micsql\ with leading LLMs and published prompt-based approaches. 

\begin{itemize}
\item \textbf{GPT-4}~\cite{achiam2023gpt} is a general LLM that serves as a strong prompt-based baseline for SQL generation.

\item \textbf{DIN-SQL}~\cite{pourreza2023din} decomposes complex questions into sub-questions to improve SQL generation accuracy.

\item \textbf{DAIL-SQL}~\cite{gao2023text} leverages in-context learning by dynamically selecting relevant demonstration examples.

\item \textbf{MAC-SQL}~\cite{wang2025mac} adopts a modular prompting strategy that divides schema understanding, reasoning, and SQL generation into multiple stages.

\item \textbf{CHESS}~\cite{talaei2024chess} employs a multi-agent LLM framework with specialized components for retrieval, schema pruning, candidate generation, and query validation, enabling efficient SQL generation.

\item \textbf{OpenSearch-SQL}~\cite{xie2025opensearch} generates SQL under a multi-agent LLM framework enhanced by an alignment mechanism, SQL-like intermediate language, and dynamic few-shot strategies.

\item \textbf{Chase-SQL}~\cite{pourreza2024chase} utilizes a multi-path reasoning framework that generates diverse SQL candidates and employs preference-optimized selection to identify the most reliable query for Text2SQL.
\end{itemize}

\subsubsection{Implementation Details}
We utilize official API services for all experiments. During the pre-processing phase, we employ the SQLite3 library in Python as the SQL executor and the TANE~\cite{huhtala1999tane} algorithm to identify functional dependencies. We use GPT-4o to generate semantic metadata and convert statistical metadata into natural language descriptions. 
During the schema linking phase, we set the retry count to 3 to ensure the SQL queries can be successfully parsed. For the SQL generation phase, we set the maximum number of interaction rounds to 6 to ensure result convergence. 
We generate 9 candidate SQL queries for each question in the BIRD dev, the BIRD test, and the Spider test sets, respectively. 
Using both metadata-complete schemas and metadata-partial schemas as inputs for schema linking, we generate 4 SQL queries based on metadata-partial schemas at temperatures \{0.1, 0.4, 0.4, 1.0\} and 5 SQL queries based on the metadata-complete schemas at temperatures \{0.1, 0.1, 0.4, 0.4, 1.0\}.

\begin{table}[hpbt]
    \centering
    \caption{Execution accuracy (EX) on the BIRD dev set and test set. All results are from publicly available papers and the official leaderboard.}
    \setlength{\tabcolsep}{0.6cm} 
    \begin{tabular}{lcc}
        \toprule
        \textbf{Method} &\textbf{Dev} &\textbf{Test} \\
        \midrule
         GPT-4 &46.35 &54.89 \\
         DIN-SQL+GPT-4 &50.72 &55.90 \\
         DAIL-SQL+GPT-4 &54.76 &57.41 \\
         MAC-SQL+GPT-4 &54.76 &57.41 \\
         MCS-SQL+GPT-4 &63.36 &65.45 \\
         CHESS+Gemini-1.5-pro &68.31 &66.53 \\
         OpenSearch-SQL &69.30 &72.28 \\
         OpenSearch-SQL w/o vote &67.80 &- \\
         Chase-SQL+Gemini &\textbf{74.90} &76.02  \\
         \textbf{\micsql\ } &74.45 &\textbf{76.41} \\
         \textbf{\micsql\ w/o vote} &\textbf{72.03} &- \\
        \bottomrule
    \end{tabular}
    \label{table:bird_result}
\end{table}
\footnotetext{\url{https://bird-bench.github.io/}}
\subsection{Main Result}
Table~\ref{table:bird_result} presents the results on the BIRD dataset, where \micsql\ achieves 74.45\% EX on the dev set and 76.41\% EX on the private test set.
To the best of our knowledge, this represents state-of-the-art performance among all published methods.
Although Chase-SQL scores slightly higher on the dev set, it relies on an ensemble of three strategies to generate 21 candidate SQL queries, which consumes more computational resources than our approach.
Moreover, the performance of our method can be further improved by increasing the number of candidates (see Section \ref{sec:parameter}). 
More importantly, we emphasize that without applying majority voting, our method significantly outperforms existing methods that rely on single-SQL generation and any single strategy employed by Chase-SQL on the dev set. 
This advantage not only demonstrates superior accuracy but also comes with significantly reduced inference time and token usage, thereby enhancing the practicality of \micsql\ for real-world applications.

We further validate the generalizability of \micsql\ on the Spider dataset. We only adjust the pluggable SQL alignment phase. This adjustment adapts our method to the output style of Spider and mitigates potential underestimation issues.
Table~\ref{table:spider_result} demonstrates that our method achieves an execution accuracy of 88.7\% on the Spider test set. Since the test set is publicly available, all baseline results are taken from published papers. This competitive performance underscores the strong generalization capability of our approach.

\begin{table}[!hpbt]
    \centering
    \caption{Execution accuracy (EX) on the Spider test set.}
    \setlength{\tabcolsep}{0.32cm}
    \begin{tabular}{lcc}
        \toprule
        \textbf{Method} & \textbf{EX} & \textbf{Training with Spider}\\
        \midrule
         GPT-4 &83.9  & \XSolidBrush \\
         MAC-SQL+GPT-4 &82.6 & \Checkmark \\
         DIN-SQL+GPT-4 &85.3 & \Checkmark \\
         DAIL-SQL+GPT-4 &86.6 & \Checkmark \\
         MCS-SQL+GPT-4 &89.6 & \Checkmark \\
         CHESS+Gemini-1.5-pro &87.2 & \XSolidBrush \\
         OpenSearch-SQL &87.1 & \XSolidBrush\\
         Chase-SQL+Gemini &87.6 & \XSolidBrush \\
         \textbf{\micsql\ } & \textbf{88.7} & \XSolidBrush\\
         \textbf{\micsql\ w/o vote} &\textbf{87.6} & \XSolidBrush\\
        \bottomrule
    \end{tabular}
    \label{table:spider_result}
\end{table}

\subsection{Ablation Study}
To better understand the contributions of individual phases in \micsql, we conduct an ablation study. Specifically, we evaluate the impact of each phase on the overall performance. The results on the BIRD dev set are summarized in Table~\ref{table:ablation}.
The baseline schema linking phase includes column example values related to the question.
The SQL generation phase is an inherent component of the framework and cannot be removed. It includes the few-shot and CoT settings. The SQL alignment phase consists of multiple components and therefore does not have an initial standalone configuration.
Table~\ref{table:ablation} reveals that: (1) All modules exhibit clear and cumulative contributions to the overall performance. This indicates that the modules are well designed to be complementary and can work effectively together within our framework. (2) Metadata-complete contexts improve EX\textsubscript{single} and EX\textsubscript{vote} by 2.80\% and 2.29\%, respectively, indicating that complete database metadata plays a crucial role in the Text2SQL task. (3) The intermediate correction mechanism improves EX\textsubscript{single} and EX\textsubscript{vote} by 1.69\% and 1.10\%, respectively, indicating that it effectively corrects errors during the SQL generation process. (4) Function alignment and output alignment make progressive performance improvements, and the rule-generated alignment strategy effectively ensures consistency between phases.

\definecolor{headergray}{RGB}{240,240,240}
\begin{table}[!hpbt]
    \centering
    \caption{Effects of individual phases on \micsql\ performance of the BIRD dev set. EX\textsubscript{single} denotes the execution accuracy when generating a single SQL, while EX\textsubscript{vote} represents the execution accuracy obtained by voting over 9 generated SQLs. `+' denotes the addition of a phase to the previous setting. Values in parentheses indicate the performance change with respect to the preceding configuration.}
    \setlength{\tabcolsep}{0.14cm}
    \begin{tabular}{lcc}
        \toprule
        \textbf{Setup} & \textbf{EX\textsubscript{single}} & \textbf{EX\textsubscript{vote}} \\
        \midrule
         GPT-4o &58.67 &60.04 \\
        \midrule
        \rowcolor{headergray}
         \textbf{+ Schema Linking} & 66.56\textsubscript{(7.87 $\uparrow$)} & 68.83\textsubscript{(8.79$\uparrow$)} \\
         \quad+ Metadata-complete &69.36\textsubscript{(2.80$\uparrow$)} &71.12\textsubscript{(2.29$\uparrow$)} \\
         \midrule
         \rowcolor{headergray}
          \textbf{SQL Generation} &- &- \\
         \quad+ Intermediate Correction &71.05\textsubscript{(1.69$\uparrow$)} &72.87\textsubscript{(1.75$\uparrow$)} \\
         \midrule
         \rowcolor{headergray}
          \textbf{+ SQL Alignment} &- &- \\
         \quad+ Function Alignment &71.18\textsubscript{(0.13$\uparrow$)} &73.20\textsubscript{(0.33$\uparrow$)} \\
         \quad+ Output Alignment &71.57\textsubscript{(0.35$\uparrow$)} &73.92\textsubscript{(0.72$\uparrow$)}\\
         \quad+ Rule-Generated Alignment &72.03\textsubscript{(0.56$\uparrow$)} &74.45\textsubscript{(0.53$\uparrow$)}\\
         \midrule
         \textbf{\micsql\ } &\textbf{72.03} &\textbf{74.45} \\
        \bottomrule
    \end{tabular}
    \label{table:ablation}
\end{table}

\subsection{Performance Breakdown}
\begin{figure}[!hptb]
	\centering
	\includegraphics[width=\linewidth]{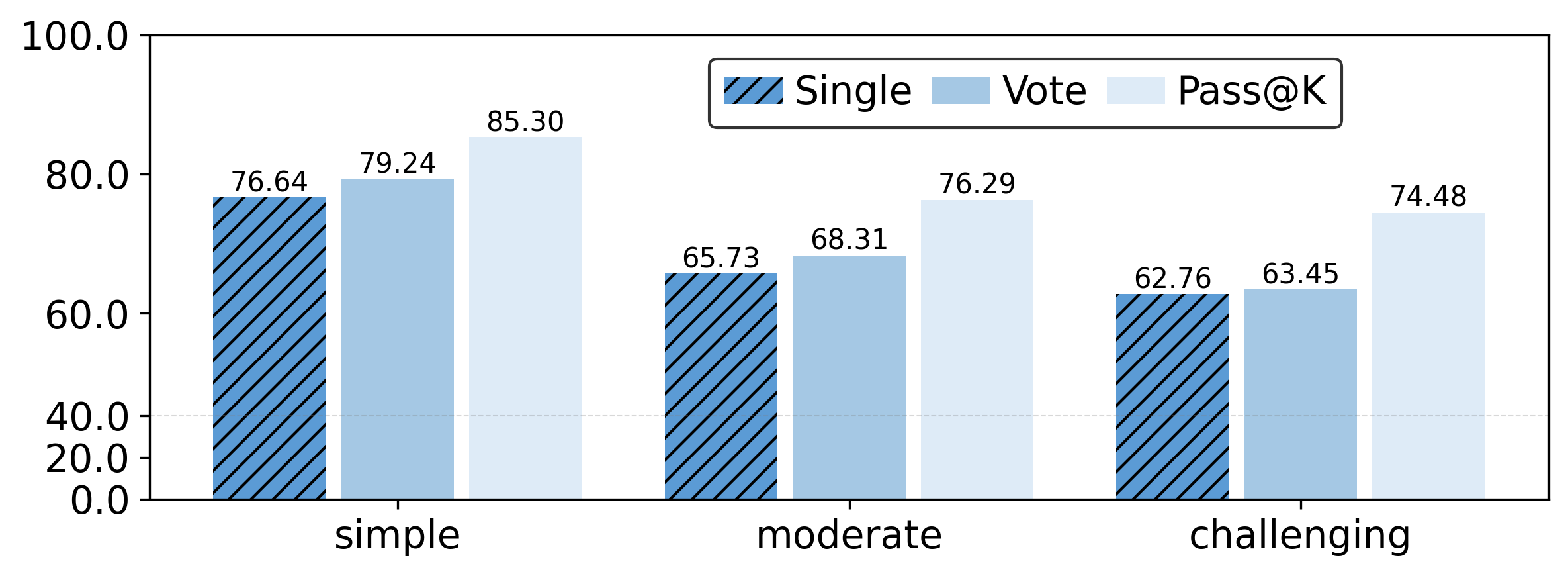}
	\caption {Execution accuracy under different question difficulties.} 
\label{pic:exp_difficult}
\end{figure}

We present a performance breakdown of \micsql\ across different question difficulty levels. This analysis provides insights into how the model behaves under varying levels of question complexity.
Figure~\ref{pic:exp_difficult} shows that model performance declines gradually across simple, moderate, and challenging difficulty levels. This trend remains consistent across single, vote, and pass@k metrics. This indicates that question difficulty significantly impacts Text2SQL.
\micsql\ achieves the highest single accuracy and upper bound performance in the simple difficulty setting. 
Furthermore, the performance gain from the voting strategy diminishes as the question difficulty increases. 
This occurs mainly because the model exhibits higher uncertainty when answering complex questions.
Therefore, employing specific selection models (e.g., Chase-SQL~\cite{pourreza2024chase} and Agentar-Scale-SQL~\cite{wang2025agentar}) can improve performance.

\subsection{Execution Cost}
We further analyze the cost of generating metadata for a single database and producing an SQL query, in terms of execution time and token consumption.
Note that different databases have different schema structures, and questions vary in complexity, which leads to variations in computational cost during inference.
In addition, the execution time is also affected by external API response time and SQL execution time.
As shown in Table~\ref{table:exp_cost}, the construction of database metadata is conducted offline, and its cost can be effectively amortized through reuse.
During metadata construction, the pattern metadata accounts for the largest proportion of token consumption, mainly because the database contains a large number of columns that require natural language pattern descriptions.
Generating example metadata is the most time-consuming step, as this module requiresbuilding indexes over all values in the database.
We refer to the online process that generates SQL queries for user questions as the generation pipeline, which takes approximately 92 seconds and costs about 0.1 dollars.
The SQL generation phase dominates both execution time and token consumption since it involves multiple interactions with the SQL executor, which introduces additional computational and communication overhead.
\begin{table}[!hpbt]
    \centering
    \caption{Execution cost of metadata construction and SQL generation pipeline.}
    \setlength{\tabcolsep}{0.14cm}
    \begin{tabular}{l|ccc}
        \toprule
        \textbf{Modular} &\textbf{Time (s)} &\textbf{Output Token} &\textbf{Cost (\$)} \\
        \midrule
         \textbf{Metadata-complete} &3700 &22000 &0.231 \\
        \midrule
         table description &130 &3500 &0.036 \\
         example  &1800 &0 &0 \\
         pattern &160 &14000 &0.144 \\
         range &70 &0 &0 \\
         semantic similarity &40 &500 &0.006 \\ 
         dependency &1500 &4000 &0.045 \\
        \midrule
         \textbf{Pipeline} &92 &7700 &0.100 \\
        \midrule
         Schema Linking &4 &300 &0.017 \\  
         SQL Generation &80 &6700 &0.072 \\
         SQL Function Alignment &4 &400 &0.006 \\
         SQL Output Alignment &4 &300 &0.005 \\
        \midrule
         SQL Selection &0.05 &0 &0 \\
        \bottomrule
    \end{tabular}
    \label{table:exp_cost}
\end{table}

\section{Analysis}
\subsection{Detail Analysis of Metadata-Complete Context}
To evaluate the effectiveness of the metadata used in this paper, we analyze various potential metadata configurations. 
Theoretically, richer metadata facilitates a deeper understanding of database structure and semantics, which leads to improved performance. 
However, excessive metadata does not necessarily enhance understanding due to the limitations of current LLMs.
Previous studies have verified the benefits of metadata, such as data examples~\cite{chen2024open, zhang2023act} and column meanings~\cite{qu2024before}. 
Therefore, we include this metadata by default in all experimental settings. 
We conduct experiments on the challenging BIRD MiniDev dataset to avoid the high computational costs of evaluating the full dataset.
Furthermore, since the alignment phase in \micsql\ does not involve the database schema, we restrict our evaluation to the schema linking and SQL generation phases.

As shown in Table~\ref{table:exp_metadata}, under the metadata-complete schema setting, the schema linking stage achieves the highest column recall, which is significantly outperforming other configurations and provides a reliable candidate set for subsequent SQL generation.
We further analyze the contributions of different metadata types and find that removing the data pattern and data range causes the largest drop across all metrics. This indicates that these two types of metadata are crucial for both column filtering and SQL generation.
When semantic similarity is removed, recall decreases while precision and F1 increase. However, execution accuracy still drops, indicating that semantic similarity plays an important role in generating correct SQL queries.
Removing dependencies and table descriptions also results in a decrease in execution accuracy during the SQL generation phase.
Based on the selected metadata-complete configuration, adding rows, null values, and size metadata causes negligible change in column filtering metrics, whereas execution accuracy decreases to varying extents.
This suggests that, given the current model capabilities, additional metadata does not effectively improve SQL generation.
With further improvements of LLMs, we expect that more comprehensive metadata can be better leveraged to bridge the semantic gap between language models and databases.
\begin{table}[!hpbt]
    \centering
    \caption{Precision, Recall, F1 of column filtering in the schema linking phase and execution accuracies (EX) in the SQL generation phase on the BIRD Minidev set. `$-$' / `$+$' indicates that the corresponding metadata is removed/added from the metadata-complete configuration.}
    \setlength{\tabcolsep}{0.22cm} 
    \begin{tabular}{l|ccc|c}
        \toprule
        \textbf{Method} &\textbf{Precision} &\textbf{Recall} &\textbf{F1} &\textbf{EX} \\
        \midrule
         \textbf{Metadata-Complete} &83.60 &96.17 &89.45 &67.5  \\
        \midrule
         - table description &83.20 &94.72 &88.59 &66.5 \\
         - pattern \& range &82.70 &94.59 &88.25 &62.0 \\
         - semantic similarity &96.12 &94.72 &95.42 &66.0 \\ 
         - dependency &83.60 &96.17 &89.45 &67.0 \\
        \midrule
         \textbf{Metadata-Complete} &83.60 &96.17 &89.45 &67.5  \\
        \midrule
         + rows &83.33 &94.99 &88.78 &66.0 \\
         + null value &83.64 &95.12 &89.01 &65.0 \\
         + size &82.93 &94.85 &88.49 &66.5 \\
        \bottomrule
    \end{tabular}
    \label{table:exp_metadata}
\end{table}

\subsection{Detail Analysis of Intermediate Correction SQL Generation}
\begin{figure}[!t]
	\centering
	\includegraphics[width=\linewidth]{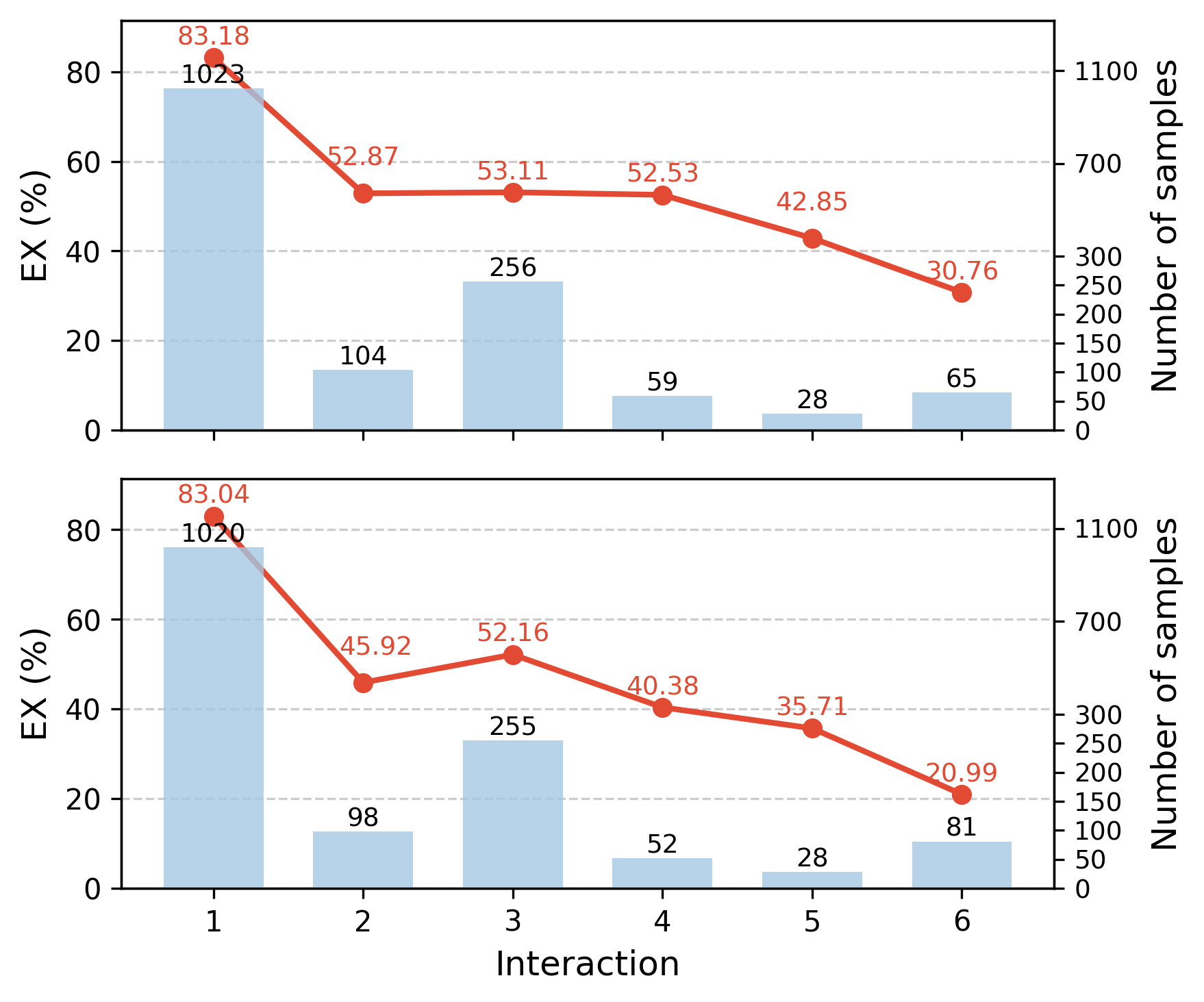}
	\caption {Execution accuracy and sample distribution across different interaction counts for the proposed intermediate correction SQL generation phase shown on the top, and the baseline method with all-at-once prompt injection shown on the bottom.} 
\label{pic:exp_generation}
\end{figure}

We further conduct a detailed analysis of the performance improvements introduced by the proposed intermediate correction SQL generation phase.
As a comparison baseline, we input all instruction prompts into the LLM simultaneously while keeping all other experimental settings unchanged.
In this setting, the LLM is required to identify relevant instructions and perform error correction after receiving SQL execution results.
Figure~\ref{pic:exp_generation} illustrates the interaction process between the model and the SQL execution tool during inference.
When the interaction count is one, the LLM determines that the draft SQL generated during the schema linking stage is sufficient to correctly answer the user query, and there is no need to further polish it.
Therefore, the LLM only invokes the SQL executor once to run the draft SQL.
This scenario accounts for the largest proportion of examples and also achieves the highest execution accuracy.
Cases with an interaction count of three occur relatively frequently.
This is mainly because such questions typically require two effective interactions during the decomposition and correction process.
One interaction is used to obtain the execution results of the sub-SQL, and the other interaction is used to generate the final SQL query.
In both settings, the execution accuracy consistently decreases as the number of interactions increases.
This trend indicates that questions requiring more interactions are generally more difficult to solve successfully.
Experimental results show that the intermediate correction strategy significantly improves execution accuracy and reduces the total number of interactions required for successful SQL generation.

\subsection{Hyperparameters in SQL Selection}
\label{sec:parameter}
\begin{figure}[!t] 
	\centering
	\includegraphics[width=\linewidth]{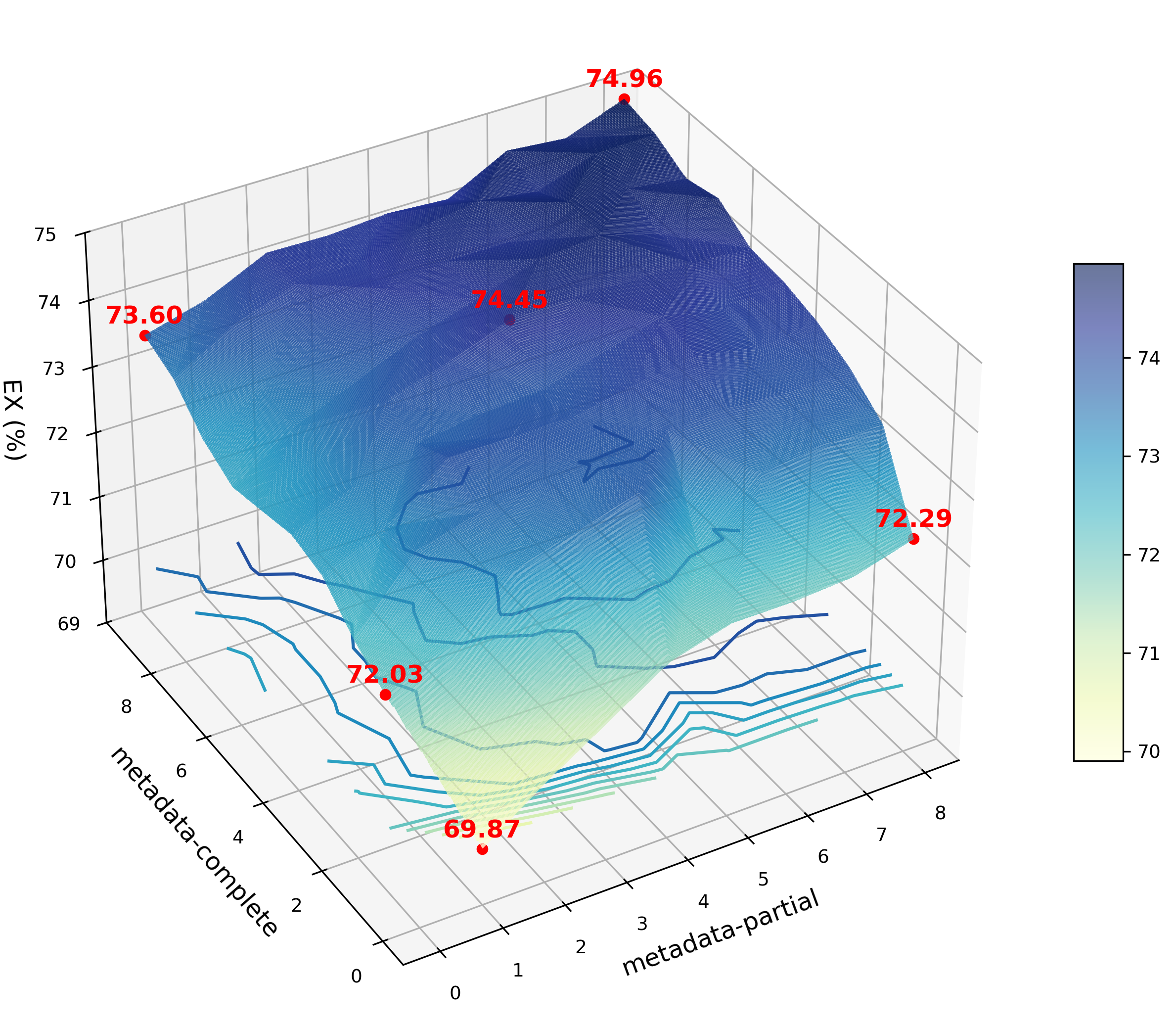}
	\caption {Impact of metadata combinations on vote performance.} 
\label{pic:exp_vote}
\end{figure}

In the SQL selection phase, we generate candidate SQL queries using schemas with two different metadata granularities and select the final SQL through a voting strategy.
To analyze the impact of metadata granularity and the number of candidates on voting, we conduct a grid search over different combinations of metadata-complete and metadata-partial schemas. As shown in Figure~\ref{pic:exp_vote},
when metadata-complete schemas are absent, the voting performance remains relatively low.
As the number of candidate SQLs generated from metadata-partial schemas increases, EX improves gradually, but the overall gain is still limited.
Specifically, the EX increases from 69.87\% to 72.29\%, which is consistently lower than configurations that include metadata-complete schemas.
When metadata-partial schemas are absent, the voting performance relies solely on metadata-complete schemas.
As the number of candidate SQL queries from metadata-complete schemas increases, EX improves from 72.03\% to 73.60\%, indicating that metadata-complete schemas provide effective support for the voting process.
When both metadata-complete schemas and metadata-partial schemas are incorporated, the results are significantly better than those obtained using only a single level of metadata.
As the number of candidate SQLs from both metadata levels increases simultaneously, the EX continues to improve and reaches a peak of 74.96\% with 8 candidates from metadata-partial schemas and 9 from metadata-complete schemas.
Overall, the experimental results demonstrate that high-quality and diverse candidate SQLs are crucial for improving EX.
As the number of candidate SQLs increases, the voting process better exploits complementary information from different levels of metadata granularity, leading to more accurate SQL selection.

\section{BIRD-clear and Result}
During the evaluation period, we find that the BIRD dev set contains a noticeable number of annotation errors.
Although the BIRD team has acknowledged this issue and plans to correct it, the revised dataset has not yet been publicly released.
Therefore, to obtain a more accurate evaluation, we manually verify and correct the BIRD dev set and construct a revised version named BIRD-clear.
Specifically, we recruit 10 graduate students with research backgrounds in databases.
The entire process lasts 15 days and consists of three stages, namely individual annotation, cross review, and final inspection. Ultimately, we have corrected 412 samples.

Each sample in the BIRD dataset contains \textit{question ID, question, evidence, and SQL.}
We determine whether the question–evidence–SQL triplet is consistent and categorize errors into three types:
\begin{itemize}
\item \textbf{Incorrect SQL}: The question and evidence clearly express a valid intent, but the SQL does not correctly implement it; therefore, only the SQL needs to be corrected.
\item \textbf{Ambiguous question}: The question description is vague or ambiguous, and either the question or the evidence needs to be modified to align with the SQL.
\item \textbf{Unanswerable question}: The question cannot be answered using the current database. For example, when the question refers to fields that do not exist, we replace the question with the closest answerable alternative, and the corresponding evidence and SQL are updated accordingly.
\end{itemize}
Note that, to simulate realistic retrieval noise, except for revisions made to resolve ambiguity, we do not add or remove hints in the evidence, even when such hints are noisy or misleading.
In addition, when the original SQL and the revised SQL produce identical execution results, we keep the original SQL even if the revised one is more reasonable, because the Text2SQL task evaluates correctness based on execution results rather than SQL semantics.

\definecolor{headergray}{RGB}{240,240,240}
\definecolor{sqlbg}{RGB}{248,248,248}
\definecolor{keywordblue}{RGB}{0,102,204}
\definecolor{keywordpurple}{RGB}{128,0,128}
\definecolor{keywordyellow}{RGB}{204,153,0}  
\definecolor{keywordpinklight}{RGB}{219,112,147}  
\newcolumntype{C}[1]{>{\raggedright\arraybackslash}m{#1}}

\begin{table*}[htbp]
\centering
\small
\caption{Error types and examples}
\label{table:exp_error}
\begin{tabular}{C{2cm} C{3.2cm} C{5.5cm} C{4.5cm} C{0.8cm}}
\toprule
\rowcolor{headergray}
\textbf{Error phase} & \textbf{Question} & \textbf{Ground Truth SQL} & \textbf{Predicted SQL} & \textbf{Ratio} \\
\midrule
\textbf{Schema Linking}
&\textbf{\textcolor{keywordyellow}{\hl{How often}}} does account number 3 request an account statement to be released? What was the aim of debiting 3539 in total?
& 
SELECT \textbf{\textcolor{keywordyellow}{\hl{T1.frequency}}}, T2.k\_symbol FROM account AS T1
INNER JOIN (SELECT account\_id, k\_symbol, SUM(amount) AS total\_amount
FROM `order` GROUP BY account\_id, k\_symbol) AS T2
ON T1.account\_id = T2.account\_id
WHERE T1.account\_id = 3 AND T2.total\_amount = 3539
&SELECT \textbf{\textcolor{keywordyellow}{\hl{COUNT(*) AS frequency}}}, k\_symbol FROM trans WHERE account\_id = 3 AND type = 'VYDAJ' GROUP BY k\_symbol HAVING SUM(amount) = 3539
& 26.88\% \\

\midrule
\textbf{SQL Generation}
&How many \textbf{\textcolor{keywordblue}{\hl[blue!15]{connections}}} does the atom 19 have?
& 
SELECT COUNT(T.bond\_id) FROM connected AS T WHERE \textbf{\textcolor{keywordblue}{\hl[blue!15]{SUBSTR(T.atom\_id, -2) = '19'}}}
& SELECT COUNT(*) FROM connected WHERE \textbf{\textcolor{keywordblue}{\hl[blue!15]{atom\_id LIKE 'TR\%\_19'}}} OR \textbf{\textcolor{keywordblue}{\hl[blue!15]{atom\_id2 LIKE 'TR\%\_19'}}}
& 40.86\% \\

\midrule
\textbf{\makecell[l]{SQL Function \\ Alignment}}
&Which country is the \textbf{\textcolor{keywordpinklight}{\hl[pink!35]{oldest}}} driver from?
& 
SELECT nationality FROM drivers WHERE \textbf{\textcolor{keywordpinklight}{\hl[pink!35]{dob}}} \textbf{\textcolor{keywordpinklight}{\hl[pink!35]{IS NOT NULL}}} ORDER BY dob ASC LIMIT 1
& SELECT nationality FROM drivers ORDER BY dob ASC LIMIT 1
& 12.90\% \\

\midrule
\textbf{\makecell[l]{SQL Output \\ Alignment}}
&What is the highest \textbf{\textcolor{keywordpurple}{\hl[purple!15]{total}}} \textbf{\textcolor{keywordpurple}{\hl[purple!15]{bilirubin level}}} recorded? List out the patient details with ID, sex and birthday with that index.
& SELECT \textbf{\textcolor{keywordpurple}{\hl[purple!15]{T2.`T-BIL`}}}, T1.ID, T1.SEX, T1.Birthday FROM Patient AS T1 INNER JOIN Laboratory AS T2 ON T1.ID = T2.ID ORDER BY T2.`T-BIL` DESC LIMIT 1
&SELECT P.ID, P.SEX, P.Birthday, L.`T-BIL` FROM Laboratory L JOIN Patient P ON L.ID = P.ID WHERE L.`T-BIL` = (SELECT MAX(`T-BIL`) FROM L)
& 19.35\% \\
\bottomrule
\end{tabular}
\end{table*}

\begin{figure}[!hptb]
	\centering
	\includegraphics[width=\linewidth]{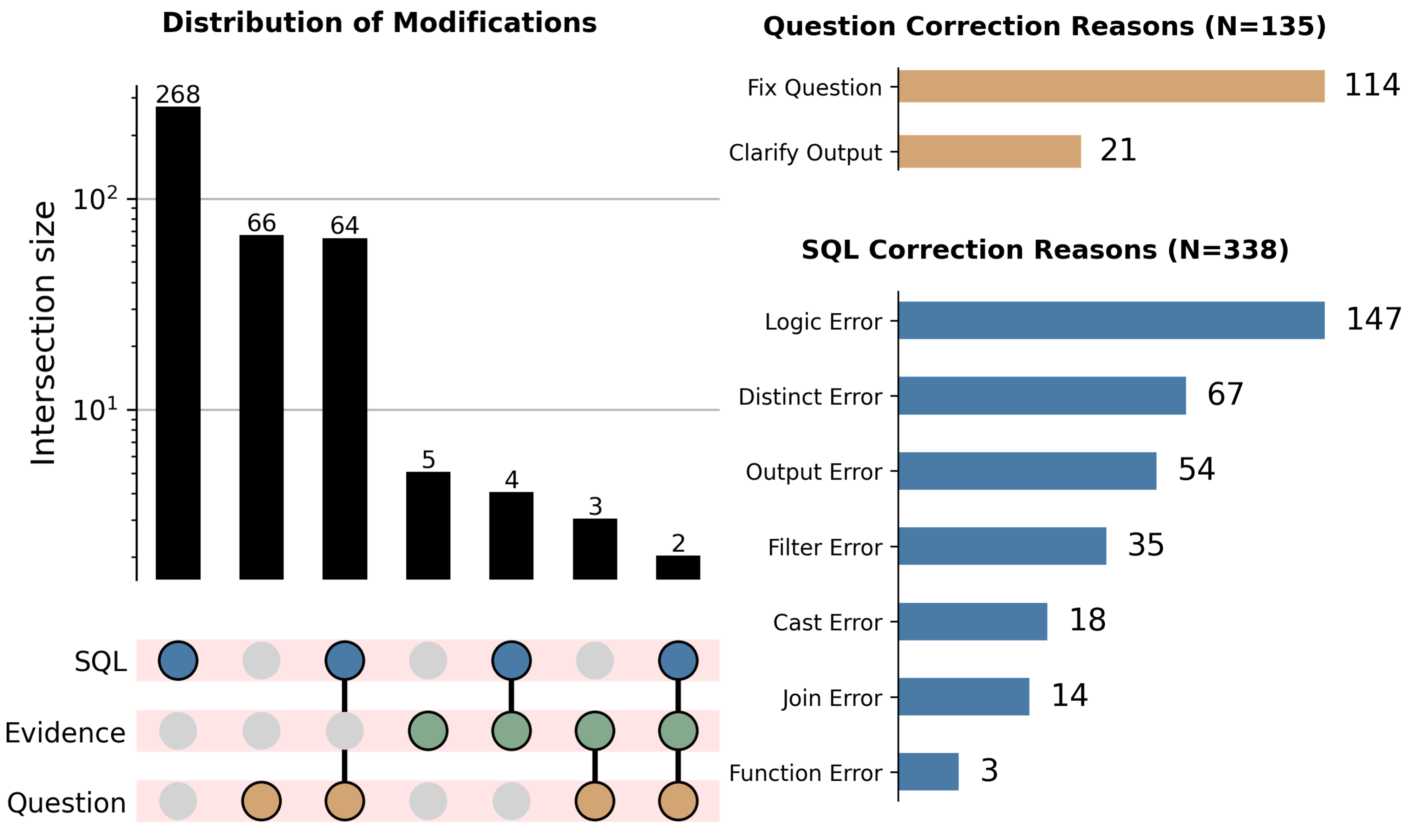}
	\caption {Distribution of modifications in the BIRD dev set. The left subfigure depicts seven possible combinations of modifications across questions, SQLs, and evidence, with gray circles marking unchanged categories. Horizontal bars indicate the total number of samples modified in each category, and vertical bars indicate the number of samples per combination. The right subfigure illustrates the distribution of question and SQL error samples.} 
\label{pic:exp_modification}
\end{figure}
Figure~\ref{pic:exp_modification} illustrates the detailed distribution of modifications.
We observe that samples with only SQL corrections account for the largest proportion, which indicates that generating accurate SQL from a clear question remains challenging, and even human annotators are susceptible to errors.
Samples requiring only question corrections typically stem from ambiguous wording.
Samples requiring both question and SQL corrections mostly appear when the SQL captures the general intent but contains minor errors.
For example, the SQL may use \textit{COUNT(id)} instead of the more appropriate \textit{COUNT(DISTINCT id)}, and the corresponding question description is also insufficiently clear.
In addition, we further categorize the specific error types for both question and SQL corrections.
For SQL corrections, the most common error type is logical errors, where the SQL semantics do not match the question.
\textit{Join Error} indicates incorrectly joined extra tables or missing required tables.
\textit{Function Error} indicates the use of inappropriate functions, such as using \textit{IIF} instead of \textit{CASE}.

As shown in Table~\ref{table:notation}, 
we partition the BIRD-clear dataset into two subsets: $\text{BIRD}_\text{c}$-412 and its complement $\overline{\text{BIRD}_\text{c}\text{-412}}$, where $\text{BIRD}_\text{c}$-412 denotes the 412 corrected samples. 
\begin{table}[hpbt]
    \centering
    \caption{Notations of dataset subsets.}
    \setlength{\tabcolsep}{0.2cm} 
    \renewcommand{\arraystretch}{1.2} 
    \begin{tabular}{|>{\centering\arraybackslash}m{1.4cm}|m{6cm}|}
        \hline
        \textbf{Notation} 
        & \textbf{Definition and Description} \\
        \hline
        $\text{BIRD}_\text{c}$-412 & 412 corrected samples in BIRD-clear\\ 
        \hline
        $\text{BIRD}_\text{o}$-412 &  corresponding 412 original samples in BIRD \\
        \hline
        $\overline{\text{BIRD}_\text{c}\text{-412}}$ & Remaining samples in BIRD-clear excluding the 412 samples \\
        \hline
    \end{tabular}
    \label{table:notation}
\end{table}
Table~\ref{table:exp_bird_clear} reports the experimental results of \micsql\ on the BIRD-clear dataset under single-SQL generation setting.
We observe that the performance of \micsql\ on $\text{BIRD}_\text{o}$-412 is significantly lower than that on $\overline{\text{BIRD}_\text{c}\text{-412}}$, and the execution accuracy on the $\text{BIRD}_\text{c}$-412 shows a clear improvement compared with $\text{BIRD}_\text{o}$-412.
These results indicate that annotation errors in the original dataset lead to an underestimation of model performance.
We also evaluate the current state-of-the-art open-source model, Arctic-Text2SQL-R1~\cite{yao2025arctic}, which is trained on large-scale SQL data using reinforcement learning.
The experimental results confirm that this underestimation phenomenon exists across different methods, while our method still achieves better performance than existing approaches on the corrected dataset.
\begin{table}[tbhp]
    \centering
    \caption{EX of different methods on BIRD-clear.}
    \setlength{\tabcolsep}{0.08cm} 
    \begin{tabular}{l|ccc|c}
        \toprule
        \textbf{Method} &\textbf{$\overline{\text{BIRD}_\text{c}\text{-412}}$} &\textbf{$\text{BIRD}_\text{o}$-412} &\textbf{$\text{BIRD}_\text{c}$-412} 
        &\textbf{BIRD-clear}\\
        \midrule
         Arctic-SQL\footnotemark &78.52 &41.99 &43.93 &69.23 \\
         \micsql\ &82.08 &44.66 &65.78 &77.70 \\
        \bottomrule
    \end{tabular}
    \label{table:exp_bird_clear}
\end{table}
\footnotetext{Arctic-SQL denotes Arctic-Text2SQL-R1-7B, the only open-source model in the Arctic-Text2SQL-R1 series.}

\subsection{Error Analysis}
To evaluate the performance of \micsql\ and reveal its limitations, we conduct a manual error analysis for each phase using the BIRD-clear dataset. Table~\ref{table:exp_error} presents the error distribution and representative examples, where each error is attributed to the earliest phase that caused the erroneous SQL query.
In the schema linking phase, errors mostly result from the failure to identify relevant columns, indicating that LLMs still have room for improvement in database understanding. 
The SQL generation phase exhibits the highest error rate, primarily due to an insufficient understanding of business logic, which leads to incorrect filter conditions or semantic errors. 
This suggests that the logical reasoning capabilities of current models remain limited. 
In the SQL function alignment phase, errors generally arise from failures to correct SQL details, whereas in the SQL output alignment phase, errors typically occur when the model selects incorrect columns required by the user.
Overall, while \micsql\ achieves significant performance improvements, certain errors inevitably persist. The error patterns across different phases reveal the limitations of LLMs in database comprehension, business logic reasoning, and detailed SQL generation.

\section{Conclusion}
Text2SQL remains challenging due to errors in database understanding and SQL generation. To address these issues, we propose \micsql, which leverages metadata-complete column contexts and an intermediate correction mechanism to facilitate performance. 
Our framework achieves superior execution accuracy without relying on massive candidate generation, thereby reducing both token consumption and latency. Experiments on the BIRD and the refined BIRD-clear benchmark show that \micsql\ outperforms state-of-the-art methods, demonstrating its efficiency and practical applicability for real-world scenarios.
\clearpage

\bibliographystyle{ACM-Reference-Format}
\bibliography{sample}

\end{document}